\documentclass{sig-alternate-05-2015}
\pdfoutput=1

\usepackage{graphicx}
\usepackage{cite}
\usepackage{microtype}
\usepackage{url}
\usepackage{listings}
\usepackage{flushend}
\usepackage{oz}
\usepackage[ruled,vlined]{algorithm2e}
\usepackage{multirow}
\usepackage[usenames,dvipsnames,table]{xcolor}
\usepackage{booktabs}
\usepackage{paralist}

\usepackage{boxedminipage}
\newenvironment{result}%
{\medskip
\noindent
\let\emph=\textbf
\begin{boxedminipage}{\columnwidth}\begin{center}\em}%
{\end{center}\end{boxedminipage}%
}

\SetCommentSty{mycommfont}

\definecolor{rltred}{rgb}{0.5,0,0}
\definecolor{rltgreen}{rgb}{0,0.5,0}
\definecolor{rltblue}{rgb}{0,0,0.5}
\definecolor{DarkGreen}{rgb}{0.00,0.60,0.00}
\definecolor{ScarletRed}{rgb}{0.80,0.00,0.00}
\definecolor{blizzardblue}{rgb}{0.67, 0.9, 0.93}
\definecolor{green-yellow}{rgb}{0.68, 1.0, 0.18}

\definecolor{dkgreen}{rgb}{0,0.6,0}
\definecolor{gray}{rgb}{0.5,0.5,0.5}
\definecolor{mauve}{rgb}{0.58,0,0.82}
\definecolor{lightgrey}{rgb}{0.90,0.90,0.90}

\definecolor{grey}{gray}{0.75}
\definecolor{light-gray}{gray}{0.80}

\newcommand\JSONnumbervaluestyle{\color{blue}}
\newcommand\JSONstringvaluestyle{\color{red}}

\newif\ifcolonfoundonthisline

\makeatletter

\lstdefinestyle{json}
{
  showstringspaces    = false,
  keywords            = {false,true},
  alsoletter          = 0123456789.,
  morestring          = [s]{"}{"},
  stringstyle         = \ifcolonfoundonthisline\JSONstringvaluestyle\fi,
  MoreSelectCharTable =%
    \lst@DefSaveDef{`:}\colon@json{\processColon@json},
}

\newcommand\processColon@json{%
  \colon@json%
  \ifnum\lst@mode=\lst@Pmode%
    \global\colonfoundonthislinetrue%
  \fi
}

\lst@AddToHook{Output}{%
  \ifcolonfoundonthisline%
    \ifnum\lst@mode=\lst@Pmode%
      \def\lst@thestyle{\JSONnumbervaluestyle}%
    \fi
  \fi
  \lsthk@DetectKeywords%
}

\lst@AddToHook{EOL}%
  {\global\colonfoundonthislinefalse}

\makeatother

\begin{document}

\definecolor{bg}{rgb}{0.95,0.95,0.95}
\newtheorem{definition}{Definition}

\title{Selecting Effective Black-box Tests With Uncertainty Sampling}
\title{Generating Effective Black-box Tests With Uncertainty Sampling}
\title{Effective Black-box Test Data Generation \\ By Uncertainty Sampling}
\title{Black-box Test Data Generation \\ By Uncertainty Sampling}
\title{Exploiting Uncertainty to Generate Effective Black-box Tests}
\title{Using Uncertainty to Guide Black-box Test Generation}
\title{Uncertainty Guided Black-box Test Generation}
\title{Uncertainty-Driven Black-Box Testing}
\title{Uncertainty-Driven Black-Box Test Data Generation}

\numberofauthors{2}
\author{
\alignauthor
Neil Walkinshaw\\
\affaddr{University of Leicester}\\
\affaddr{Leicester, UK}
\alignauthor
Gordon Fraser\\
\affaddr{University of Sheffield}\\
\affaddr{Sheffield, UK}
}

\maketitle
\begin{abstract}
We can never be \emph{certain} that a software system is correct simply by testing it, but with every additional successful test we become less \emph{uncertain} about its correctness. In absence of source code or elaborate specifications and models, tests are usually generated or chosen randomly. However, rather than randomly choosing tests, it would be preferable to choose those tests that decrease our uncertainty about correctness the most. In order to guide test generation, we apply what is referred to in Machine Learning as ``Query Strategy Framework'': We infer a behavioural model of the system under test and select those tests which the inferred model is ``least certain'' about. Running these tests on the system under test thus directly targets those parts about which tests so far have failed to inform the model. We provide an implementation that uses a genetic programming engine for model inference in order to enable an uncertainty sampling technique known as ``query by committee'', and evaluate it on eight subject systems from the Apache Commons Math framework and JodaTime. The results indicate that test generation using uncertainty sampling outperforms conventional and Adaptive Random Testing.

%
\end{abstract}




\definecolor{bg}{rgb}{0.95,0.95,0.95}

\section{Introduction}

Testing software components without access to source code or hand-crafted models is challenging because there is no guidance for the selection of test inputs. Selection is invariably guided by intuition or, if automated, by random or quasi-random input generation algorithms \cite{Hamlet1994,Chen2004,Claessen2011}. Left to chance alone, random test sets can easily fail to expose facets of software behaviour that depend upon specific input characteristics. Furthermore it can become exceedingly difficult to reason about the adequacy of a randomly-generated test set, especially for non-numerical programs without an operational profile \cite{Hamlet1994}.

Recently, several ``Learning-Based Testing'' (LBT) techniques have emerged \cite{Raffelt2009,Walkinshaw2009,Lei2013,Fraser2015} that aim to address these limitations. LBT techniques are based on the idea, first espoused by Weyuker \cite{Weyuker1983} and Budd and Angluin \cite{Budd1982}, that there is a natural duality between inductive model inference and software testing. The former seeks to infer a general model of  behaviour for a system from an incomplete sample of observations of its behaviour. The latter seeks to identify the smallest possible set of observations that are required to expose the full range of behaviour. Although the ultimate purposes are different, both are bound by an intrinsic challenge: attempting to establish the link between the often infinite range of externally observable behaviour of a system and a finite sample of observations (or vice versa). 

LBT techniques seek to exploit this duality by using Machine Learning algorithms to infer input / output models from test executions. These models can then be used to derive new test cases. The rationale is that this ought to form a virtuous loop (or, to adopt Popper's terminology, a cycle of ``conjecture and refutation'' \cite{Popper1953}) where the inferred models become increasingly detailed and accurate, and thereby drive the test generation to produce increasingly rigorous test sets. 

The step of generating new test inputs from an inferred model is especially important. New test inputs ought ideally to expose `new' aspects of software behaviour that have not featured in previous test executions. Intuitively, the test generation approach tends to be closely tied to the type of inferred model (e.g., if an approach infers a state machine, it will tend to adopt a state machine testing algorithm to derive new tests \cite{Walkinshaw2009,Meinke2011}). 

Unfortunately,  there are two barriers that currently restrict LBT approaches to relatively specific classes of relatively small-scale software systems:

\begin{compactenum}
\item The dependence between the type of inferred model and the test generation approach can be highly limiting. Whole families of powerful Machine Learning algorithms have to be excluded as they do not produce explicit, `testable' models. 
\item The application of model-based test generation approaches to inferred models can yield large numbers of test cases, which hampers scalability. Many of the generated tests are of little \emph{utility} to the learner. Whereas the goal is to find `counter-examples' to the inferred models, the majority of test cases can merely end up corroborating what is already known. 
\end{compactenum}

In this paper we investigate the possibility of using an Active Learning query strategy frameworks \cite{Settles2010,Seung1992} to circumvent these limitations. In Machine Learning, query strategy frameworks provide a means by which to use an existing inferred model (or set of models) to select further samples that are most likely to be of ``high utility'' to the learner -- i.e. provide information that is not already contained within the training set. These tend to be based on the principle that the best samples are those whose prediction elicits the highest degree of \emph{uncertainty} with respect to the current model. In the context of LBT, if one accepts the existence of a relationship between the adequacy of a test set and the accuracy of a model inferred from it, then it should follow that test cases selected by an effective uncertainty sampling technique should form an effective basis for test case selection.  

In detail, the contributions of this paper are as follows:
\begin{compactitem}
\item We introduce the first application of query strategy frameworks to test generation (Section~\ref{sec:qbc_test}).
\item We present an implementation of a query strategy framework for test generation using Query By Committee~\cite{Seung1992} on inferred models (Section~\ref{sec:qbc_test}).
\item We propose the use of Genetic Programming~\cite{Koza1992} as a basis for model inference, as it directly enables Query By Committee (Section~\ref{sec:qbc_test}).
\item We present an implementation of an LBT-based testing using query strategy frameworks, based on Genetic Programming and Query By Committee (Section~\ref{sec:qbc_test}).
\item We present an empirical evaluation on eight functions provided within the Apache Commons Math and JodaTime frameworks, using mutation testing to assess the effectiveness of the generated test cases (Section~\ref{sec:eval}).
\end{compactitem}

Our experiments demonstrate that uncertainty sampling leads to a higher mean number of mutants detected than random or adaptive random testing (the baseline techniques we use in this paper). It also tends to require fewer test executions to produce higher numbers of mutants. This is especially valuable for test-scenarios where there is a non-trivial cost associated with test execution (e.g. tests take a prohibitive amount of time, or their outputs need to be checked by a human test-oracle).

\section{Automated Black-box Testing}

Black-box testing in general refers to the concept of testing a software system without access to its source code. Ideally, black-box testing is driven by detailed formal specifications or test models, which enable techniques to automatically generate tests, and act as test oracle that decides whether a given test execution revealed a fault or not. In practice, such specifications are not always available, in which case automated generation of tests is limited to few options.

\subsection{Random Testing}

The most common approach to test automation in the absence of formal specifications and source code is to randomly select tests, for example using a uniform distribution on the input space or an operational profile \cite{Hamlet1994}. The effectiveness of random testing highly depends on the specifics of the system under test: Random testing is generally unlikely to find specific input values~\cite{Arcuri2012}, and may perform poorly at covering the underlying behaviour of the program.

Adaptive Random Testing (ART)~\cite{Chen2004} aims to alleviate these problems by ensuring that tests are spread out across the program input space as much as possible. In general, ART works iteratively by repeatedly sampling a set of random inputs, and out of this set selecting the input that is most different to previously executed tests as the next test to run on the system under test. While there is evidence that this approach makes the selected tests more effective than a completely random selection, every test input adds to complexity of generating the next test input, because there is an additional point in euclidean space against which to measure the next group of random inputs. 

If running a test on a system under test is cheap, then pure random testing may be more effective than ART~\cite{Arcuri2011} as it can simply execute significantly more tests in the same time as ART. However, in practice test execution can often take a long time, and the absence of an automated oracle (e.g., a formal specification) may make it necessary to manually investigate every single test outcome. Thus, we assume that it is desirable to generate the most effective set of tests, rather than relying on the ability to run large sets of potentially redundant tests.

\subsection{Learning-Based Testing}

We use the term `Learning-Based Testing' (LBT) to refer to the (now relatively broad) family of techniques that seek to use Machine Learning to support the generation of test cases. The idea was first explored by Weyuker \cite{Weyuker1983} and Budd and Angluin \cite{Budd1982} in the early eighties. For the subsequent 15 years it was the subject of some predominantly theoretical research \cite{Cherniavsky1987,Zhu1992,Romanik1997}. However, over the subsequent 15 years it adopted a more practical bent, with several authors developing accompanying proof-of-concept tools \cite{Bergadano1996,Raffelt2009,Lei2013,Walkinshaw2009,Meinke2011,Fraser2015,Papadopoulos2015}.

\addtolength{\textfloatsep}{-\baselineskip}
\begin{algorithm}[t]
\SetKwInput{KwUses}{Uses}
\KwIn{$SUT$,$TestInputs$}
\KwUses{$terminate$, $execute$, $selectInputs$, $inferModel$}
\KwResult{$TestInputs$}
$hyp$ $\leftarrow \emptyset$ \;
$Executions$ $\leftarrow \emptyset$ \;
\For{($input \leftarrow TestInputs$)}{
	$Executions \leftarrow Executions \cup execute(input)$\;
}
\While{($\neg$ terminate($Executions$,$hyp$,$SUT$))}{
  $hyp$ $\leftarrow$ inferModel($Executions$)\;
  $NewInputs$ $\leftarrow$ selectInputs($hyp$,$SUT$)\;
  $Executions \leftarrow Executions \cup execute(SUT,NewInputs)$\;
  $TestInputs \leftarrow TestInputs \cup NewInputs$\;
}
\Return{$TestInputs$}\;

\caption{Generic LBT procedure}
\label{alg:iterativeAlgorithm}
\end{algorithm}

\addtolength{\textfloatsep}{1\baselineskip}

Algorithm \ref{alg:iterativeAlgorithm} shows the main generic LBT steps: 

\begin{compactitem}
\item The algorithm starts with an initial set $TestInputs$ of inputs to the program. This may be empty, but it may also be an established test set that we wish to improve.
\item The loop of model inference and test generation is executed until a stopping criterion \textbf{terminate($Executions$,$hyp$)} evaluates to true. For example, it might attempt to establish the equivalence between the inferred model $hyp$ and the system under test $SUT$, and return true if the model is sufficiently similar in some sense \cite{Raffelt2009}. It might alternatively simply terminate after a fixed number of iterations, if $Tests$ reaches a particular size, or there has been no change to $hyp$ after a certain number of iterations.  
 \item In this loop, the first step is to infer a predictive input/output model \textbf{hyp} of the program using the function \textbf{inferModel($Executions$)}. The type of the model can vary, and depends on the nature of the system under test. Proposed techniques have adopted state machines \cite{Raffelt2009,Walkinshaw2009}, decision trees \cite{Briand2009,Fraser2015} and Daikon invariants~\cite{Ghani2008}.
 \item The input to the \textbf{inferModel} function are the executions, i.e., the input/output pairs resulting from executing the test inputs $TestInputs$ on the system under test $SUT$ using function \textbf{execute($SUT$,$Inputs$)}.
 \item Finally, \textbf{selectInputs($hyp$,$SUT$)} selects new inputs. The test generation strategy might be random \cite{Papadopoulos2015}, driven by source code coverage~\cite{Fraser2015}, or using a model-based test  algorithm with respect to  $Mod$ \cite{Raffelt2009}.
\end{compactitem}

Much of the research on combining inference and testing has focussed on the interplay between the \emph{terminate} and \emph{inferModel} functions --- on the ability to leverage inference mechanisms to provide more meaningful adequacy criteria. This is what motivated most of the early research into the area as well~\cite{Weyuker1983,Cherniavsky1987,Zhu1992}. Recent inference and test generation techniques have been combined to guarantee that the behaviour of the system has been exercised to a certain extent. For example, several researchers have combined Angluin's $L*$ inference technique~\cite{Angluin1987} with established state machine testing techniques~\cite{Raffelt2009,Lei2013} and showed that these lead to strong guarantees that the inferred model accurately represents what has been explored. 

\begin{figure}[t]
	\includegraphics[width=\columnwidth]{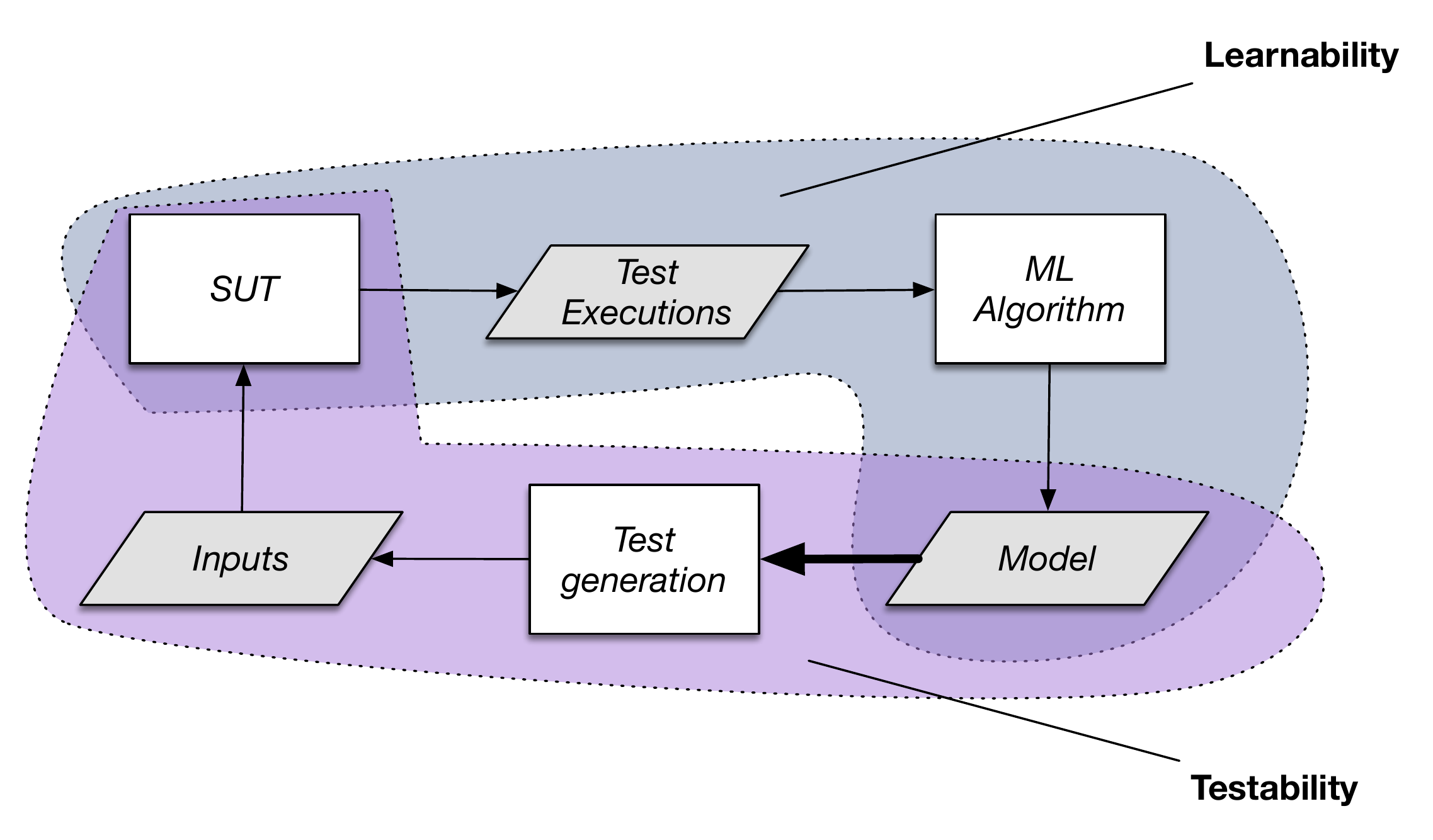}
    \vspace{-2em}
	\caption{The relationship between `learnability' and `testability' in LBT.}
	\label{fig:problem}
    \vspace{-1em}
\end{figure}

\subsection{Limitations of Learning-Based Testing}


LBT techniques tend to be limited in their practical applicability because they rely on the inference of models that not only approximate the behaviour of the SUT, but are \emph{also} usable as a basis for automated test generation. As illustrated in Figure \ref{fig:problem}, the processes of inference and testing are highly interdependent; the model has to be \emph{learnable} from the SUT \cite{Valiant1984}, but also has to be \emph{testable}, in the sense that it can provide a suitable basis for the generation of new test cases. This explains why LBT techniques so far have been largely restricted to state machines \cite{Raffelt2009,Walkinshaw2009,Lei2013}, decision trees \cite{Briand2009,Papadopoulos2015}, and invariants~\cite{Ghani2008}.  As a consequence, entire families of Machine Learning algorithms that infer models that are harder to subject to symbolic reasoning are excluded, even if they could potentially infer more accurate models from a broader class of SUT's. These include kernel-method techniques such as SVM \cite{Vapnik1995}, Neural Net learners \cite{Bishop1995}, and Genetic Programming~\cite{Koza1992}. 

Aside from the constraints mentioned above, the use of `off-the-shelf' test generation techniques, coupled with the iterative nature of the approach, can lead to scalability problems. Test generation algorithms generate test cases without considering the test cases that have been generated in previous iterations. This can produce very large test sets, especially if the testing algorithm in question produces large numbers of test cases anyway (e.g., the popular  W-Method \cite{Chow1978} for state machines is a good example of an algorithm that does not scale well, and has often used for test-driven model inference \cite{Raffelt2009}). 

The primary challenge addressed in this paper is to find a means by which to remove the constraints on the classes of Machine Learners that can be applied to LBT. Although we want to remove the constraints on the types of models that are inferred, it is crucial that they can still guide test generation, and do so in a scalable way that does not needlessly re-test behaviour that has already been tested.

\section{Query Strategy Frameworks} 
\label{sec:uncertainty_sampling}

In this paper we will show how the above problem can be solved by the use of query strategy frameworks, a core facet of active learning techniques \cite{Settles2010} in Machine Learning. Precisely how this is achieved will be described in the next section. Here we provide a generic introduction to query strategy frameworks and Query by Committee.

\subsection{Query Strategy Frameworks}

Machine Learning algorithms can be broadly categorised into conventional (\emph{passive}) Machine Learners and \emph{active} Machine Learners. Passive learners infer models from data that is given to them before learning commences. Active learners might start from some given data, but crucially are also imbued with the ability to obtain further data. The learner might surmise that the inferred model is incomplete because the initial sample lacked data of a certain characteristic, so the learner can set out to obtain more relevant data, which it can use to refine its model. 

The active learning setting gives rise to the \emph{query strategy problem} \cite{Settles2010}. The process of obtaining a sample might be expensive, so it is consequently important to keep the number of queries (additional samples) down to a minimum. However, any additional data that \emph{is} sampled must be of a \emph{high utility} --- i.e., it must lead to improvements in the model inferred by the learner. This problem has been the subject of a large amount of research over the past two decades (a good overview is provided by Settles \cite{Settles2010}). The essential goal is to avoid selecting a query that fails to add new information that is of value to the learner. Any new data should ideally confound the predictions of the current model.

One factor that plays a key role in selecting queries is the notion of \emph{uncertainty}. Given a data-point that was not part of the original training set (referred to as a `query'), the degree of uncertainty exhibited by the current hypothesis model as to how it should be classified can provide an indication of how useful it would be to obtain a real sample. The goal is thus to identify queries for which the level of confidence in the corresponding output is at its lowest, with the aim of eliciting aspects of behaviour that were perhaps under-represented in the training sample. 

One key challenge is to find a suitable metric that can be used to assess this ``uncertainty'' for a given model prediction. For statistical Machine Learners, where the output is often in the form of a probability distribution, numerous uncertainty sampling techniques have been developed \cite{Settles2010}. However, in the context of LBT, models such as inferred state machines tend not to be probabilistic.

\subsection{Query By Committee}

There is a `trick' that enables the application of uncertainty sampling even when the inferred models are themselves not probabilistic. If one can, from a given sample, infer \emph{multiple} different models, then it becomes possible to use their mutual agreement / disagreement to estimate a level of uncertainty and use this as a basis for uncertainty sampling  \cite{Seung1992}. 
Algorithm \ref{alg:qbc} shows the Query By Committee (QBC) approach proposed by Seung \emph{et al.} \cite{Seung1992}. 

\begin{algorithm}[t]
\SetKwInput{KwUses}{Uses}
\KwIn{$Train$, $i$, $s$,$comitteeSize$,$randomPoolSize$}
\KwUses{$learnMultiple$,$best$, $computeUtility$, $randomPoints$}
\KwResult{$Hyp$}
$Hyp$ $\leftarrow \emptyset$ \;
\For{$i$ iterations}{
	$Hyp \leftarrow learnMultiple(Train,comitteeSize)$\;
	$U \leftarrow randomPoints(randomPoolSize)$\;
	\For{$s$ iterations}{
		\tcp{Pick a point $u \in U$ with max utility}
		$u = argmax_{x \in U}||computeUtility(Hyp,x)||$\;
		$l = label(u)$\;
		$Train \leftarrow Train \cup \{l\}$\;
		$U \leftarrow U \setminus \{u\}$\;
	}
}
$Hyp \leftarrow learnMultiple(Train)$\;
\Return{$best(Hyp)$}\;
\caption{Query By Committee}
\label{alg:qbc}
\end{algorithm}

\begin{compactitem}
\item The entire process iterates for a fixed number of iterations~($i$).
\item At each iteration, the \textbf{learnMultiple} function produces a ``committee'' of hypothesis models. This is conventionally achieved by Ensemble Methods \cite{Melville2004}, which produce different hypotheses by inferring models from different samples of the training set (in this paper we will illustrate an alternative approach of using the population generated by a Genetic Programming algorithm). 
\item Once the models have been inferred, the \textbf{randomPoints} function generates a set of random `inputs' $U$ -- in Machine Learning terms this is a set of \emph{unlabelled} data points. The size of $U$ is determined by the $randomPoolSize$ parameter.
\item The nested for-loop then essentially picks a subset of $s$ points in $U$. These are selected by evaluating each point in $U$ to determine those points about which the inferred models $Hyp$ are \emph{least} in agreement (as computed by the \textbf{computeUtility} function). In other words, these points would be of most \emph{utility} to the learner. 
\item Once these points are selected, they are labelled with the \textbf{label} function, added to the training set, and the process iterates.
\item After the final iteration, a set of models is inferred from the aggregate training set, and a model is selected to be returned by the \textbf{best} function. The selection criteria can vary depending on the inference approach -- one straightforward option (adopted in this paper) is to return the model that best predicts the outputs (or `labels') produced by $Train$.
\end{compactitem} 

There is a clear similarity between the QBC algorithm and the LBT algorithm in Algorithm \ref{alg:iterativeAlgorithm}. Both involve loops, where models are inferred at each iteration. In both cases, the models are used as a basis for selecting more data (test inputs in the testing context, unlabelled data points in the Machine Learning case). There are also two significant differences. In the case of QBC, the output is the final inferred model, whereas in LBT the output is the data that was used to infer the model (the test set with its outputs). In LBT, there is no fixed approach to generate test data -- it could be random, or adopt a model-based testing algorithm. In QBC, there is only one approach; regardless of the type of model or system, a random pool of unlabelled data points are generated, and the best $s$ points are chosen based on the `uncertainty' that they elicit from the inferred committee of models.

\section{Applying QBC to Test Generation}
\label{sec:qbc_test}

In the context of Machine Learning, QBC enables uncertainty-based sampling to occur, regardless of the type of model that is inferred. In this paper we produce the Testing By Committee approach, which applies QBC to LBT to circumvent the dependence between the model inference algorithm and the test-generation algorithm. In principle this enables LBT to use \emph{any} model inference algorithm, and to select test cases based on the combined uncertainty of the inferred models.

In this section we first set out our \emph{Test By Committee} algorithm, which combines LBT with QBC. We then provide technique that implements this approach using Genetic Programming as a basis for the model inference. 

\subsection{Test By Committee}

\begin{algorithm}[t]
\SetKwInput{KwUses}{Uses}
\KwIn{$SUT$,$Tests$,$s$,$i$,$comitteeSize$,$randomPoolSize$}
\KwUses{$execute$,$learnMultiple$, $randomInputs$,$computeUtility$}
\KwResult{$Tests$}
$Hyp$ $\leftarrow \emptyset$ \;
\For{$i$ iterations}{
 $Hyp \leftarrow learnMultiple(Tests,comitteeSize)$\;
	$U \leftarrow randomInputs(SUT,randomPoolSize)$\;
	\For{$s$ iterations}{
 		$u = argmax_{x \in U}||computeUtility(Hyp,x)||$\;
 		$l = execute(u)$\;
 		$Tests \leftarrow Tests \cup \{l\}$\;
 		$U \leftarrow U \setminus \{u\}$\;
 	}
}
\Return{$Tests$}\;
\caption{Testing By Committee}
\label{alg:qbc-test}
\end{algorithm}

Our proposed `Test By Committee' (TBC) algorithm is shown in Algorithm \ref{alg:qbc-test}. It clearly combines Algorithms \ref{alg:iterativeAlgorithm} and \ref{alg:qbc}. The key similarities and dissimilarities are as follows:
%
\begin{compactitem}
\item As with QBC, we limit the number of iterations to a fixed number $i$ (though it would certainly be possible to integrate something more elaborate, along the lines of the $terminate$ function in Algorithm \ref{alg:iterativeAlgorithm}).
\item The step of $learnMultiple$ is the same as in Algorithm \ref{alg:qbc}; a population of models are inferred using either ensemble methods or, as we will demonstrate, population-based learners such as Genetic Programming.
\item To generate the candidate test inputs, we introduce a new function \textbf{randominputs}. The $SUT$ is only used to gain information about its interface. Once the types of the interface are known, inputs are formulated as combinations of random values of the appropriate types. 
\item The process of adding new tests to the test set is the same as in Algorithm \ref{alg:qbc}. For $s$ iterations, the best candidate is selected from $U$ by seeing which candidate test case causes the most disagreement amongst models in $Hyp$. The chosen test is then executed to identify its actual output, and this is then added to $Tests$ (it is also removed from $U$ to avoid re-selection).
\end{compactitem}

Many of the steps are in effect the same as they are in conventional LBT. However, two steps are very different, and therefore require a more in-depth discussion. The model inference step ($learnMultiple$) requires multiple models. The process of selecting the best candidate test case ($computeUtility$) is also new in the context of testing, and requires more details. In both cases, there are many possible ways in which they could be implemented. In the following two subsections, we describe how we have chosen to implement them for our proof of concept.

\subsection{Learning Multiple Models by Genetic Programming}
\label{sub:gp_qbc}

To produce the models required for Query-by-Committee it is possible to use a Genetic Programming (GP) inference engine \cite{Poli2008}. A GP evolves programs of a given target language towards an optimisation goal, specified by a fitness function. As mentioned previously, in principle any inference technique could be applied (underpinned by Ensemble Methods \cite{Dietterich2000}). However, (a) the intrinsic population-based nature of GPs renders them suitable for QBC, and (b) GPs can easily be adapted to different types of languages, making them well suited for modelling programs in different domains.


For space reasons, we only provide the essential details of GPs here, and refer the reader to Poli \emph{et al.}'s GP field guide \cite{Poli2008}, along with our source code\footnote{https://bitbucket.org/nwalkinshaw/efsminferencetool}
for further details. In (tree-based) GPs, candidate programs are `evolved' as abstract syntax trees, where branch nodes correspond to `non-terminals' representing functions, and leaf-nodes represent atomic values or variables (terminals). The basic loop is as follows (details on the terms in italics will be elaborated in the next section):

\begin{compactenum}
\item Generate an initial population of programs as random compositions of non-terminals and terminals.
\item Execute each evolved program and evaluate it according to some \emph{fitness function}.
\item \emph{Select} the best programs from the population.
\item Create a new population by a process of \emph{cross-over} and \emph{mutation}.
\item Repeat from step 2 until some stopping criterion is met.
\end{compactenum}

In its traditional application, the result of the GP is the program with the best fitness value, which represents the best solution. In our case, we can exploit the population-based nature of the GP: At the end of the search, the population consists of a range of slightly varied candidate solutions optimised for the problem at hand. 

\subsection{Generating Test Cases by QBC}

The first step to applying QBC is to select the committee. For this we select the fittest set of chromosomes $Hyp$. The size of this set is determined by the parameter $committeeSize$.
The query generation step involves generating a pool of random inputs $U$, and then assessing every $u \in U$ to find the one that creates  most `uncertainty' according to the set of inferred models in $Hyp$ (in our case the set of chromosomes inferred by the GP). Every potential test input $u$ is executed on every model $h\in Hyp$, and the outputs are recorded. The input that produces the greatest spread of predictions is then chosen to be executed on the real SUT.


\section{Implementation}

We have implemented the approach described in the previous section. The implementation targets numeric programs without side-effects returning single outputs. In this section, we provide details of this implementation.

\subsection{A GP for Programs with Primitive Types}

In this section we elaborate the detailed aspects of the generic GP algorithm shown in Section~\ref{sub:gp_qbc}.

\textbf{Fitness function:} 
The fitness function provides a metric for the accuracy of the inferred program to predict the SUT. Fitness is evaluated by executing a candidate program on all existing test inputs, and comparing the outputs to those that were actually observed in the trace data. 

\textbf{Selection:} Step 3 selects good candidates from the population, so that they can be fed into the next generation. A popular approach, which we adopt here, is Tournament Selection \cite{Poli2008}. 
In our case the selection process is \emph{elitist}, this means that the best individual from one generation is always preserved for the next one.

\textbf{Crossover and Mutation:} The candidates that were selected in step 3 are subjected to a mixture of crossover and mutation (the frequency at which they occur is given in probabilistic terms). We choose to use the most common form cross-over called \emph{subtree-crossover} \cite{Poli2008}.
Mutation is carried out by selecting a random node in a tree and changing it. If the selected node happens to be a terminal, its value is simply changed. If it is a non-terminal, we replace its subtree with a randomly generated one. 

Arbitrary crossover or mutation can easily lead to nonsensical programs - for example by using String terminals with a function that expects integer parameters. Strongly-typed GP \cite{Poli2008} prevents this from happening by ensuring that every terminal and non-terminal has a declared type. 

\textbf{Termination and result:} The loop terminates once a candidate has been identified that cannot improve in terms of fitness, or once the number of iterations hits a given limit. 


\begin{table}[t]
\caption{Non-terminals and Terminals chosen for our experiments}
\label{tab:terms}
\def\arraystretch{1.5}
	\begin{tabular}{p{0.1 \columnwidth}p{0.8 \columnwidth}}\toprule
		\multicolumn{2}{l}{\textbf{Non-Terminals}}\\ \midrule
		Double (D) & add(x:D,y:D), subtract(x:D,y:D), multiply(x:D,y:D), divide(x:D,y:D), power(x:D,y:D), root(x:D, y:D), cast(x:I), if(x:B,y:D,z:D), cos(x:D), exp(x:D),log(x:D)\\
		\midrule
		Integer (I) & cast(x:D)\\
		\midrule
		Boolean (B) & and(x:B,y:B), or(x:B, y:B), LT(x:D,y:D), GT(x:D,y:D), EQ(x:B,y:B), EQArith(x:D,y:D),EQString(x:S,y:S)\\
        \midrule
		Logic (all) & if-then-else(a:B,b:D,c:D),if-then-else(a:B,b:I,c:I),if-then-else(a:B,b:S,c:S),if-then-else(a:B,b:B,c:B)\\
		\midrule
		\multicolumn{2}{l}{\textbf{Terminals}}\\ \midrule
		Double (D) & all variable names in $Vars$ of type double, one free variable limited to the interval $[-2,2]$, -1.0\\
		\midrule
		Integer (I) & all variable names in $Vars$ of type integer, one free variable limited  to the interval $[-2,2]$, 0\\
		\midrule
		Booleans (B) & All variable names in $Vars$ of type Boolean, \texttt{true}, \texttt{false}.\\ 
        \midrule
		Strings (S)& All variable names in $Vars$ of type String, any customised pre-defined String values.\\
        \bottomrule
	\end{tabular}
    \vspace{-1em}
\end{table}

\textbf{Terminals and Non-Terminals:} The choices of terminals and non-terminals are shown in Table \ref{tab:terms}. In general, of course, the choice of GP operators is flexible, and is ideally informed by knowledge about the system being inferred. In our case, we sought a reasonably general set that can be applied across a range of programs. The question of how to refine the selection of terminals and non-terminals to best suit a SUT is part of our ongoing work.

\subsection{Generating Test Cases by QBC}


For the purpose of this work (as a proof of concept), we are restricting ourselves to a particular class of system that produces single numerical outputs (either integers or numbers with decimal places). Our initial use of standard deviation proved to be problematic, as it could often produce a misleadingly high value for the situation where most of the models were in fact in agreement, but one ``rogue'' model had produced an extreme value. 

To address this problem, we instead opted for the Mean Absolute Deviation (MAD) value \cite{Kader1999}, which is less vulnerable to data-spikes. For a set of values $X=\{x_1,\ldots,x_n\}$, $MAD(X) = \frac{1}{n} \sum_{i=1}^{n}|x_i-m(X)|$, where $M(X)$ calculates the mean of $X$.

It is necessary to select a value to accommodate the situation where an inferred model returns either infinity or Not a Number (e.g., because an inferred model divides by zero), but the SUT returns a valid value. The value should be high, to indicate that the model is incorrect, but cannot be too high (e.g., Double.MAX\_VALUE), because this prevents the calculation of an accurate mean over multiple outputs. In this case, we substitute the result with a value of 10,000,000 (this was a somewhat ad-hoc choice, and establishing a more justified value is part of our future work).

\subsection{Example -- The BMI Calculator}
\label{sub:illustration}

This section contains a brief walk-through of TBC. As a SUT we choose a simple BMI calculator. This takes as input two numbers (height in meters and weight in kilograms), and returns a ``Body Mass Index'' value, calculated as $\frac{weight}{height^2}$. For our technique to operate, we do not need to be able to look at the internal implementation, but only need to know of the interface. However, to provide a complete overview, let us assume that the calculator is implemented as a bash script, with the following source code:

\begin{lstlisting}[language=Bash,mathescape=false]
#!/bin/bash
awk "BEGIN {print $2 / ($1 * $1)}"
\end{lstlisting}

Our implementation accepts a specification of the interface in the following self-explanatory JSON format. 

\begin{lstlisting}[style=json,mathescape=false]
{
	"command": "bmi.sh",
	"parameters":[
		{
			"name": "height",
			"type": "double",
            "max": "100",
            "min": "-100"
		},
		{
	        "name": "weight",
			"type": "double",
            "max": "100",
            "min": "-100"
        }
	],
	"output":[
        {
            "name": "output",
            "type": "double"
        }
	]
}	
\end{lstlisting}

Finally, we provide an existing basic test set that we wish to improve upon. 
Our implementation accepts a space-separated text file, where the order of values is taken to be the order of parameters in the specification file (height followed by weight):

\begin{lstlisting}[style=json,mathescape=false]
1.7 50
1.8 70
1.9 100
1.7 110
0.0 5
5.0 0
\end{lstlisting}

With reference to the TBC process in Algorithm \ref{alg:qbc-test}, the BMI represents the $SUT$, and the above list of test sets represents $TestInputs$. For the sake of illustration, we will only show one iteration ($i=1$), and we will only add a single test set in this iteration $s=1$. To illustrate how new test cases are selected, we set $randomPoolSize$ to 3, although this would usually be much higher (in the evaluation we will set it to 1000).

\begin{figure}[t]
	\includegraphics[width=\columnwidth]{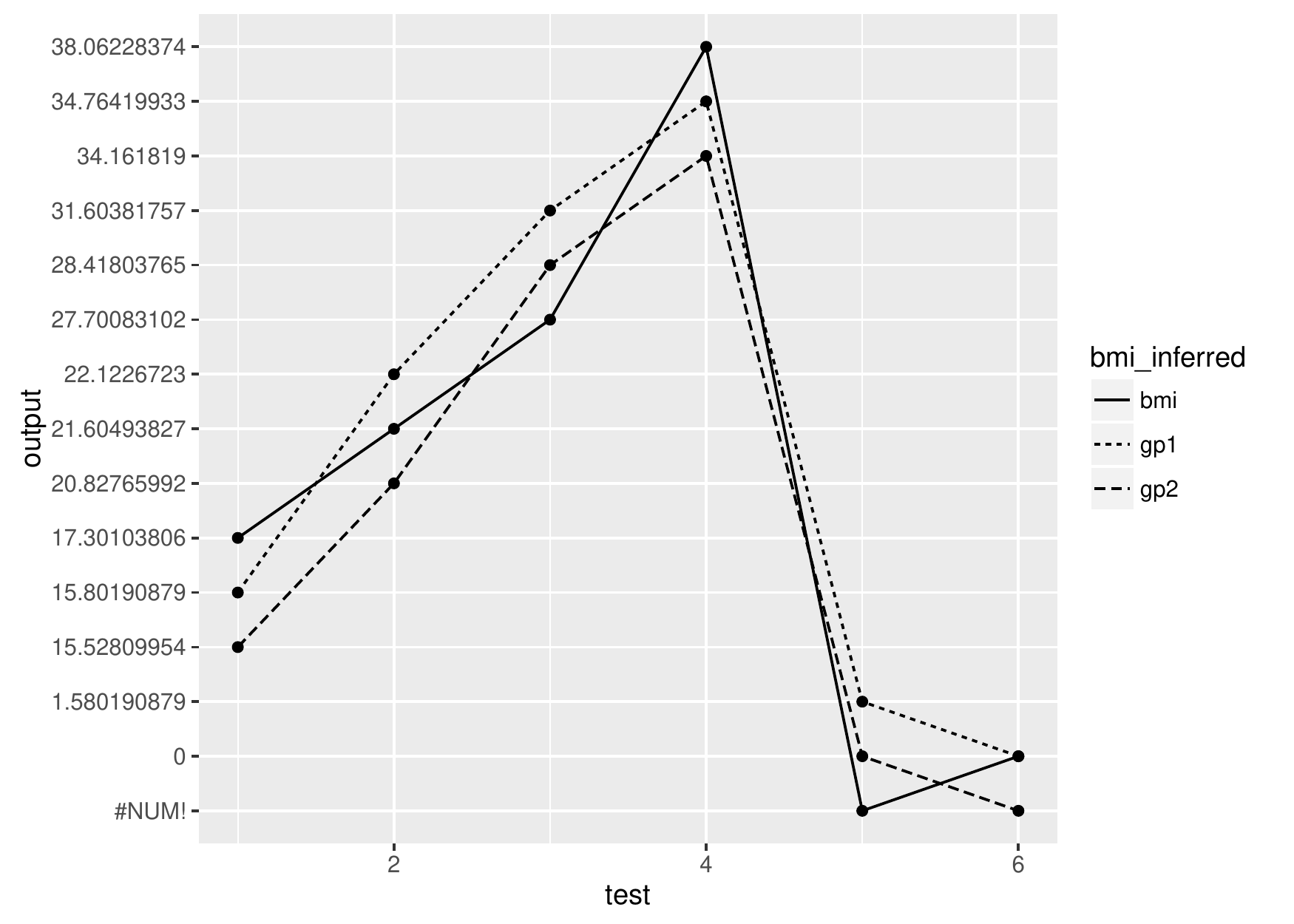}
    \vspace{-2em}
	\caption{Comparison of expected versus inferred outputs after one iteration wrt. BMI example.}
	\label{fig:gpOut}
	\vspace{-1em}
\end{figure}

The TBC algorithm begins by inferring the ``committee'' $Hyp$  via $learnMultiple$. In our case, this produces the top 10 chromosomes. To give an idea of what is inferred, two of the fittest GP programs after the first iteration is as follows:

\begin{lstlisting}[mathescape=false]
	gp1: Mult(weight,Exp(-1.1518922634307343)))
	
	gp2: Div(height,Exp(height-Log(weight))
\end{lstlisting}

Although they are clearly inaccurate, we can assume that (as the fittest members of their pool of solutions), they at least \emph{approximate} the output. This is illustrated in Figure \ref{fig:gpOut}, which plots outputs (the dashed and dotted lines) against the expected output (the plain line), for all test inputs. 

As the next step, $randominputs$ produces a set of $randomPoolSize$ inputs (in this case three). The resulting inputs are shown on the left in Table \ref{tab:inputs}. For each input, the disagreement between the models is calculated as the Maximum Average Deviation (as described previously), shown in the right-hand column. From this, it is clear that the second input produced a huge divergence between the two inferred models. 

The input with the highest MAD value is thus added to the test set, and the TBC process moves to the next iteration. This time, thanks to the new test execution, the inferred models ought to be more precise, and lead to test cases that explore new aspects of the input domain.

\begin{table}
\caption{Proposed inputs and MAD calculation}
\label{tab:inputs}
\centering
\begin{tabular}{rrr}\toprule
\textbf{height} & \textbf{weight} & \textbf{MAD}\\
\midrule
87.95 & 50.49 & 3.99\\
\textcolor{red}{-62.41}&\textcolor{red}{91.14} & \textcolor{red}{\textbf{1.80E+30}}\\
26.44 & 56.65 & 4.48\\ \bottomrule
\end{tabular}
\vspace{-1em}
\end{table}

\section{Evaluation}
\label{sec:eval}

In this section we seek to assess the effectiveness of TBC at generating rigorous test sets. Of primary concern is the question of whether TBC can detect more faults than baseline testing techniques. In this evaluation we use random testing and Adaptive Random Testing  \cite{Chen2004} as the baseline. Accordingly, the first research question is as follows: 

\vspace{0.5em}
\noindent\textbf{RQ1: Do TBC-generated test sets expose more faults than random and ART-generated test sets?}
\vspace{0.5em}

One further question is concerned with the \emph{efficiency}. In the even that TBC does not ultimately expose a larger number of errors than other techniques, it might still expose the same number of faults, but after executing fewer tests, which would render it more efficient:

\vspace{0.5em}
\noindent\textbf{RQ2: Are TBC-generated test sets more efficient at exposing faults than random and ART-generated test sets?}

\subsection{Subjects}
\label{sub:subjects}

We chose six units within the Apache Commons Math framework (version 3.6)\footnote{\url{https://commons.apache.org/proper/commons-math/}} and two units within JodaTime (version 2.9.3)\footnote{\url{http://www.joda.org/joda-time/}}. The units were chosen on the following criteria:
\begin{compactitem}
\item It must accept a single (set of) input parameters -- i.e., it must not require sequences of method calls (apart from the call to the constructor, which was included in the test-wrapper, see below).
\item It must produce a single output value.
\item The parameters accepted by the unit under test (and the output value returned by it) must either  be primitive data types that are supported by our GP implementation, or be complex objects where the constructor accepts primitive data types. 
\item The unit in question must be invoked by one of the Apache Commons Math or JodaTime test sets (so that we can obtain use these tests to infer the first model).
\end{compactitem}

\begin{table}
\caption{Subject Systems.}
\label{tab:systems}
\resizebox{\columnwidth}{!}{
\begin{tabular}{@{}lp{2.7cm}rr@{}} \toprule
\textbf{Component} & \textbf{Functionality} & \textbf{Exec. LOC}& \textbf{Tests}\\
\midrule
\texttt{\scriptsize BesselJ} & \texttt{\scriptsize value}&1,211 & 699\\

\texttt{\scriptsize Binomial} & \texttt{\scriptsize binomialCoefficientDouble}&501 & 3,000*\\ 

\texttt{\scriptsize DerivativeStructure} & \texttt{\scriptsize asinh} & 360& 3,000*\\ 

\texttt{\scriptsize Gamma} & \texttt{\scriptsize regularizedGammaQ}&783&4\\

\texttt{\scriptsize Erf} & \texttt{\scriptsize erf} & 763& 116 \\
\texttt{\scriptsize RombergIntegrator} & \texttt{\scriptsize RombergIntegrator} & 735& 4 \\
\texttt{\scriptsize Period} & \texttt{\scriptsize toStandardWeeks} & 1128& 5 \\
\texttt{\scriptsize Days} & \texttt{\scriptsize daysBetween} & 1251& 8 \\
\bottomrule
\end{tabular}
}
\vspace{-1em}
\end{table}

The eight units in Table \ref{tab:systems} represent the first  units that were encountered in each system. Where a package contained a large number of possible varieties (e.g., calculations of derivatives), we chose one at random, and avoided choosing multiple units in the same collection. Where an initial test set was particularly large (in some cases they contained $>$ 20,000 executions of the SUT), we sampled 3000 executions at random to ensure that the fitness functions in the GP could be evaluated in a reasonable amount of time.  These are marked with a `*' in Table \ref{tab:systems}. 

Apache Commons Math and JodaTime were chosen because they are written in Java, which enables us to use the Major mutation framework \cite{Just2011} and because they have a reasonably extensive set of unit tests (enabling us to use these as a starting point for the learning-based testing). Their details are shown in Table \ref{tab:systems}. The sizes of the various functionalities have to be treated as approximate. To provide the LOC of the entire libraries would be a gross overestimation. To provide the LOC for a single class would be a gross underestimation (especially in the case of Apache Commons Math, where a large portion of the functionality is contained within the very large \texttt{org.apache.commons.math4.util.FastMath} class). We provide the total LOC within the library tracked (using IntelliJ) when executing all generated test sets for a given SUT. 

\begin{table}
\centering
\caption{Mean number of mutants killed after 60 iterations. Highest values are in bold. The significance of the Mann-Whitney test is indicated in parentheses. No significance - $p>0.05$ is  (-) ,  $p<0.05$ is (*), and $p <0.001$ is (***).}
\label{tab:means}
\resizebox{\columnwidth}{!}{
\begin{tabular}{lrrlrl} \toprule
SUT & TBC            & \multicolumn{2}{c}{Random}                   & \multicolumn{2}{c}{ART}  \\
\midrule
BesselJ     &   \textbf{447.50}    & 442.83 & (***)             &442.93 & (*)            \\
Binomial  &      \textbf{30.53}   & 29.03 &(***)           & 29.20 & (***)  \\
DerivativeSin  &  \textbf{55.93}  & 51.20 &(***)           & 50.07 & (***)   \\
Erf        &    \textbf{190.52}    & 188.62 &(***)   & 189.33 & (***)    \\
 Gamma   &  \textbf{208.23}         & 206.90 &(-)  &   205.60 & (-)    \\
 Romberg Integrator &\textbf{87.77}& 87.63 &(-) & 87.46 &(-) \\ 
 periodToWeeks&\textbf{304.95}& 249.52 &(*) & 271.58 &(-) \\ 
 daysBetween&\textbf{72.13}& 50 &(-) & 49.53 &(*) \\ 
 \bottomrule
\end{tabular}
}
\vspace{-1em}
\end{table}

It is important to note that these selection criteria are in part so restrictive for the sake of control in our experiment. In practice, if we wanted to test a system for which our current GP was not sufficient, we would resort to a different Machine Learner, or add the requisite terminals and non-terminals to the GP. However, in our case, this special treatment would obviously bias the results. To avoid bias, we thus restrict ourselves to a subset of systems that are at least compatible with our choice of GP.

\subsection{Methodology}

To gauge the performance of TBC in comparison with the `state of the art', we compared the mutation scores for its test sets against randomly generated test inputs, and test sets generated by Adaptive Random Testing (ART) \cite{Chen2004}. For ART, an important factor is the choice of distance function to distinguish test sets. In our case, since most of the inputs were numerical, we chose the Euclidean distance function, which tends to be the distance measure of choice.

All of the techniques were provided with an interface specification file, which contained the various parameters, and the ranges for any numerical parameters. If parameters were strings, the potential value-selections were explicitly enumerated.
To avoid biasing results, we did not use any domain to set numerical variable range boundaries, and adopted a conservative approach;  we looked at the ranges in the given test sets, and expanding these ranges with a substantial buffer in either direction (e.g., if the range of the test cases was from 0 to 10, we would set the range from -100 to 100). The full configuration files, along with all other materials used for this experiment are available online\footnote{Omitted for double-blinding
}.

\begin{figure*}[t]
	\centering
    \setlength{\tabcolsep}{0.5em}
\begin{tabular}{cccc}
    \includegraphics[trim={0.7cm 0.5cm 0.25cm 0.1cm},clip,width=0.18\textwidth]{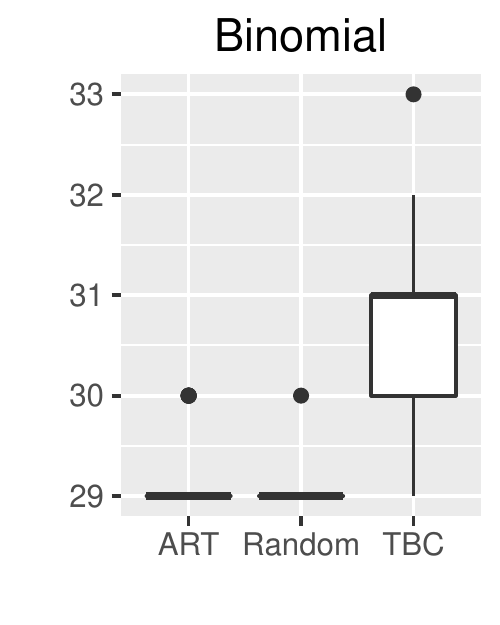} & \includegraphics[trim={0.7cm 0.5cm 0.25cm 0.1cm},clip,width=0.18\textwidth]{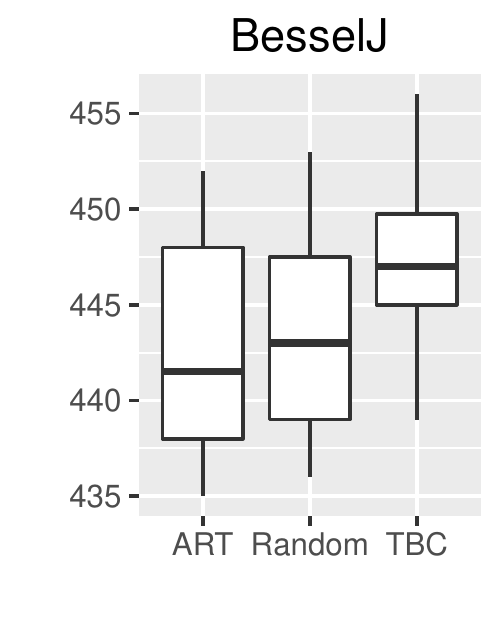}& \includegraphics[trim={0.7cm 0.5cm 0.25cm 0.1cm},clip,width=0.18\textwidth]{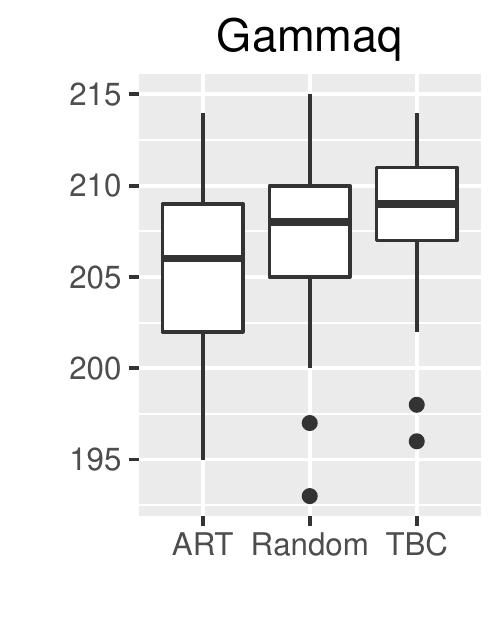}& \includegraphics[trim={0.7cm 0.5cm 0.25cm 0.1cm},clip,width=0.18\textwidth]{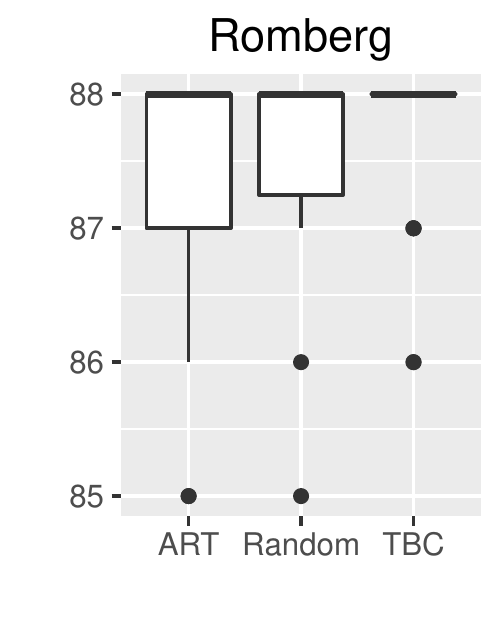}\\
     \includegraphics[trim={0.7cm 0.5cm 0.25cm 0.1cm},clip,width=0.18\textwidth]{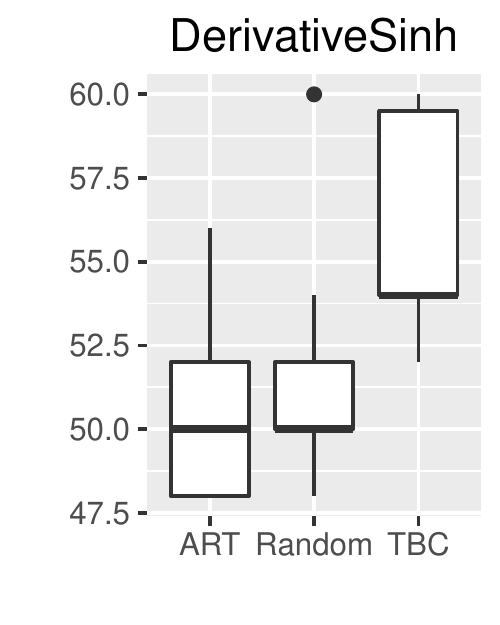} &\includegraphics[trim={0.7cm 0.5cm 0.25cm 0.1cm},clip,width=0.18\textwidth]{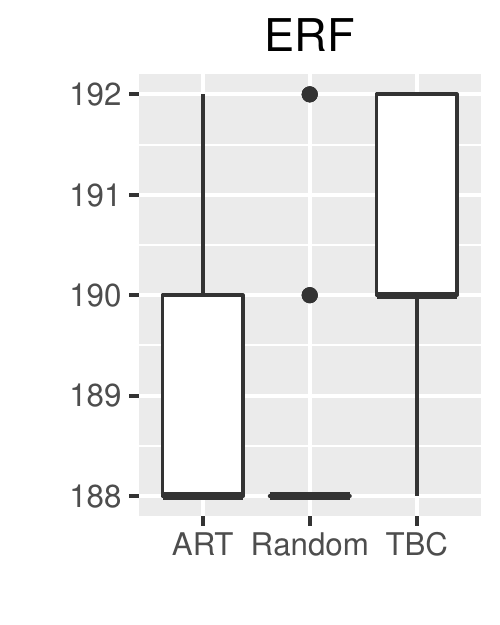}& \includegraphics[trim={0.7cm 0.5cm 0.25cm 0.1cm},clip,width=0.18\textwidth]{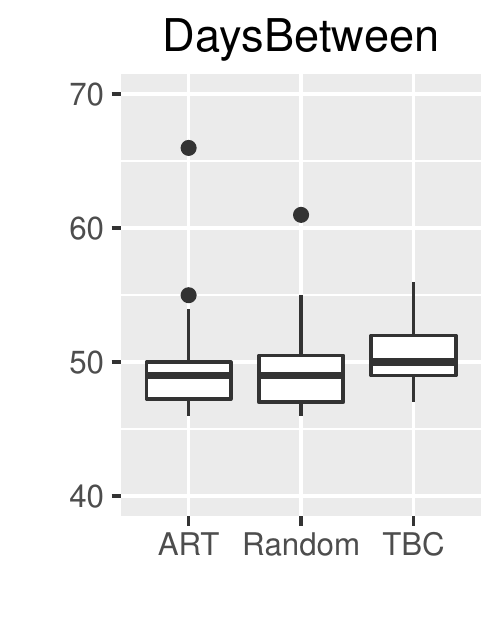} &\includegraphics[trim={0.7cm 0.5cm 0.25cm 0.1cm},clip,width=0.18\textwidth]{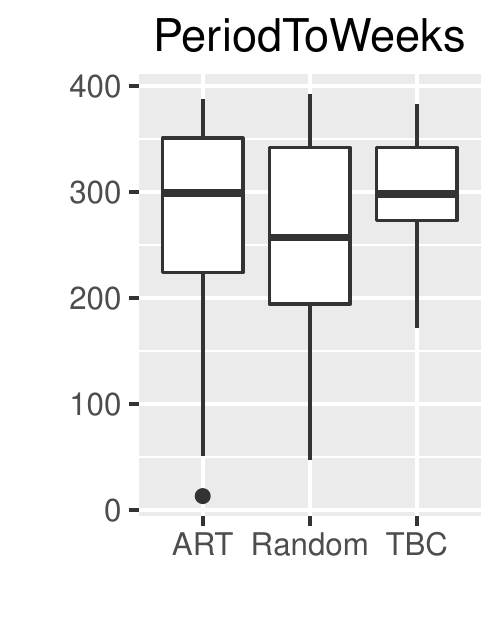}
     \end{tabular}
     \vspace{-1em}
	\caption{Mutation Scores after 60 iterations, starting from given test sets.}
	\label{fig:boxplots}
\end{figure*}

To gauge how effective a test set is at exposing faults, we employed mutation testing \cite{Jia2011}.  We used the Major Java mutation testing framework (version 1.6, with all mutants)~\cite{Just2011}. We seeded mutants conservatively, by selecting any classes that were executed by the initial set of tests (we could not seed mutants in every class in the system because of the resource constraints of mutation testing). It  does not make sense to measure the mutation score as the proportion of mutants killed, because the conservative seeding strategy will invariably mean that this proportion is liable to be very small (for example, all of the units use a fraction of the  \texttt{org.apache.commons.math4.util.FastMath} class). Instead, we simply compare the absolute numbers of mutants detected, which suffices to provide valid answers to our two research questions.

To prevent any bias arising from configurations, we used the same configuration for TBC across all experiments. For the GP configuration we used the set of terminals and non-terminals detailed in Table \ref{tab:terms}. We used a population size of 800, with a crossover-rate of 0.9, a mutation rate of 0.1, a maximum term-depth of 10 and a tournament size of 6 \cite{Poli2008}. We set the number of tests generated per iteration 1000, and the number selected for addition to the test set to 5.

To answer RQ1, we analysed the mutation scores that were computed after 60 iterations, grouped according to the technique (TBC, ART, and Random). To compare them we carried out two (non-parametric) Wilcoxon Rank Sum tests per SUT (having confirmed that the distributions are not normally distributed according to the Shapiro Wilks test). The first null-hypothesis was that the mutation scores for TBC are smaller than those for random tests. The second null-hypothesis was that  the mutation scores for TBC are smaller than those for ART tests. The distributions were also visualised as box-plots.

To answer RQ2 (how much more effective is TBC?), we recorded the last iteration at which TBC produced the highest mutation score (versus ART and Random). We also plotted the trajectories of the means to show how the trajectories differed over the course of the 60 iterations.

\begin{figure*}[t]
	\centering
	\includegraphics[width = \textwidth]{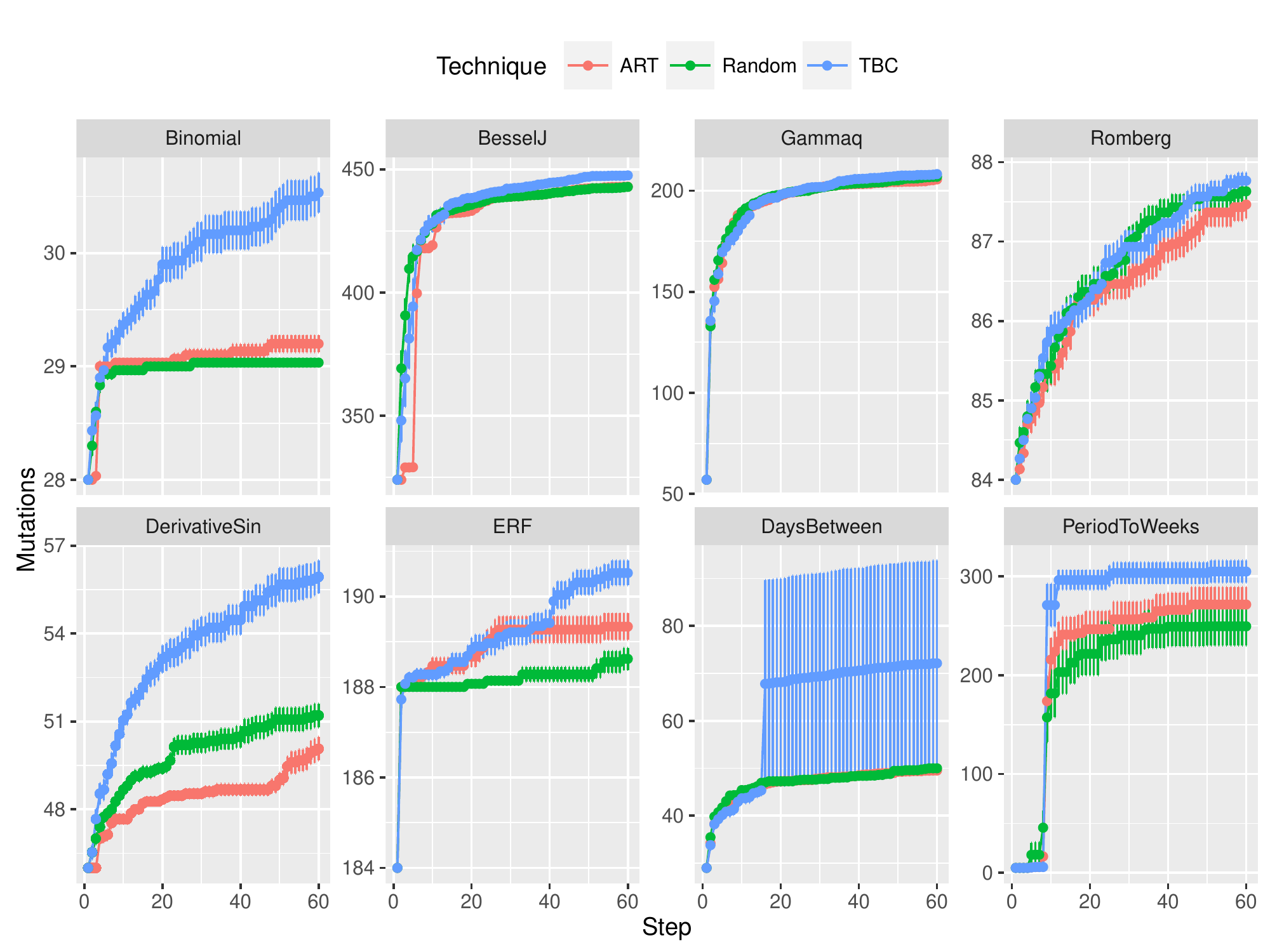}
        \vspace{-1em}
	\caption{Mutation Scores for every iteration, starting from given test sets.}
	\label{fig:trajectories}
        \vspace{-1em}
\end{figure*}

\subsection{Results for RQ1: Effectiveness}

The mean numbers of mutants killed for each system are shown in Table \ref{tab:means}. The distributions are also visualised as box-plots in Figure  \ref{fig:boxplots}. The table shows that, after 60 iterations, TBC has killed the highest mean number of mutants for every program. The improvement over ART and random testing varies substantially between the systems. For BesselJ, Binomial, Derivative Sinh, and ERF, the difference is statistically significant; this is corroborated in the box plots. In three of these systems (Binomial, Derivative Sinh and ERF), difference is so marked that the lower quartile for TBC is higher than the upper quartile for ART and Random. 

For Gamma, Romberg, PeriodToWeeks and DaysBetween although the mean is higher for TBC, the differences are not statistically significant (they are partially significant for PeriodToWeeks and DaysBetween). Looking at the box plots, in all cases apart from PeriodToWeeks the boxes for TBC are noticeably elevated. In the case of PeriodToWeeks, the median score for TBC is the same as ART (even thought the mean score is substantially higher). This is largely due to one particular execution that achieved a particularly large number of mutations. In all cases, the difference in distributions is particularly marked at the lower end; ART and Random have lower minimum scores, and lower lower-quartiles than TBC, which indicates that TBC is more consistent. 

\begin{result}
RQ1: In our experiments, TBC was  more effective than random testing and ART. In all cases there was a higher mean number of mutants killed, and the difference in distributions was significant in 4/8 SUTs.
\end{result}

\subsection{Results for RQ2: Efficiency}

We discuss the relative efficiency of TBC versus ART and random testing by looking at how rapidly TBC out-performs the other approaches (by achieving a higher mean number of mutation faults without being overtaken in subsequent iterations). Figure \ref{fig:trajectories} shows the average mutation scores and their standard deviations throughout the 60 iterations. It is important to note the differences in scales; the different SUTs give rise to markedly different numbers of mutants. This means that similar differentials in the mean numbers of mutants on different plots can appear markedly different. We discuss the various trajectories by starting with the systems where the performances are most similar.

In all of the studied systems, TBC eventually kills more mutants on average than random and ART testing. In some systems the numbers of faults detected remain similar throughout, whereas in others TBC significantly outperforms ART and Random from the start. These cases are discussed in more detail below.

As one might expect from the results for RQ1, the trajectories in the Romberg and Gammaq SUTs are visually similar; these are the systems where the relative performance between the techniques is at its closest. In the Romberg SUT, TBC is consistently better than ART from iteration 20 onwards, but only outperforms random testing after iteration 50. In Gammaq, TBC consistently outperforms ART and random from iteration 23 onwards, though only marginally. 

Perhaps more surprisingly for both JodaTime systems PeriodToWeeks and DaysBetween the trajectory for TBC is  noticeably higher than for ART and Random. For PeriodToWeeks, the number of mutants killed for TBC rapidly increases after 10 iterations to a level that  ART and Random only start to approach after 40-50 iterations.

In ERF, both ART and TBC outperform random testing from the start. ART and TBC are similar up to iteration 40, where ART continues to plateau at 189 whilst the mean number of killed mutants for TBC rises to over 190.

In Binomial, BesselJ, and DerivativeSinh, the results for TBC are markedly better from the start. In the case of BesselJ the difference may look smaller, but this is because of the scale of the graphs. In BesselJ the mean TBC score after 60 iterations is 447.5, whereas for  scores for ART and Random are approximately 443; this difference of 5 is in fact larger than the differences in the other systems.


\begin{result}
RQ2: In our experiments, TBC was significantly more efficient at exposing faults than random testing and ART.
\end{result}

\subsection{Threats to Validity}

\textbf{Threats to external validity: }
The answers to RQ1 and RQ2 can only validly be applied to systems of a similar character to those tested here. We have only tested eight systems from two frameworks. This means that they will often have shared developers, and they all deal with similar domains. We have additionally restricted ourselves to units that are functional, which do not accept sequential inputs (as discussed in Section \ref{sub:subjects}).  To attenuate this risk, we attempted to make the selection of SUTs as indiscriminate as possible within our broader selection constraints. The SUTs presented here are the first ones we encountered that fitted our criteria. However, to truly address this threat a larger study on a more diverse range of SUTs is needed, which is what we will be doing in our future work.

As mentioned previously, the choice of value ranges for the parameters is important for all of the techniques. There is a high probability that our choice of ranges is not ideal (given that we avoided using domain knowledge to avoid bias). It is possible that, for certain range limitations, the differences between the various techniques are reduced (i.e., if the value ranges are reduced). Investigating the relationship between the selection of value ranges and the relative performance of these techniques is something that we are exploring as part of our ongoing and future work.

\textbf{Threats to internal validity: }
The mutation score depends upon the seeding of mutants. It is possible that code was executed that was not seeded with mutants, thus skewing the results. We attempted to limit this possibility by tracking the execution of code with profiling tools.

\subsection{Discussion}

The results indicate that TBC tends to detect more faults with fewer tests than the baseline techniques. However, during the experiments a further factor became apparent that did not favour TBC: time. For ART 
every test input adds to complexity of generating the next test input, because there is an additional point in euclidean space against which to measure the next group of random inputs. Arcuri and Briand made this point in their critique of ART \cite{Arcuri2011}, where they showed that if time is taken into account instead of the number of tests, then ART was by some distance inferior to conventional random testing.

This question of time is even more pertinent to TBC than ART (indeed, it applies to every LBT technique). LBT involves the repeated execution of the given (and increasing) set of tests (Random and ART do not). It involves model inference, which again takes time. With the use of GP, inference time is tied to the number of available tests (since these evaluate the fitness function). For the Binomial system (the most time-consuming system studied), the full 60 iterations took on average 12 hours. For ERF (one of the least time consuming systems) it took on average 89 minutes. 

The timing question is clearly an important one to address, and a larger empirical study will be incorporating this. Looking at some of the trajectories in Figure \ref{fig:trajectories} (such as Binomial and DerivativeSin) it is doubtful whether random tests would catch up with TBC, even if we did allow for a large disparity in the number of tests. In any case, it is not necessarily always possible; some test cases simply take long to execute (e.g. if they involve complex processing or network communications), so the ability to execute huge numbers of tests rapidly is not always an option. Also, even if it is an option, the availability of many tests is not necessarily desirable either, especially if checking the outputs is a non-trivial task.

\section{Conclusions and Future Work}

In this paper we have made an explicit connection between the problems of test data generation in Software Engineering and sampling in active Machine Learning. Our solution proposes the use of uncertainty sampling as a means by which to generate suitable test data. We have provided a proof of concept implementation, along with the results of an empirical study of eight units within the Apache Commons Math and JodaTime frameworks. The initial results are encouraging. Our TBC approach outperforms random testing and Adaptive Random Testing.

Although promising, the approach has also given rise to several important questions, which were touched upon in the discussion of the threats to validity for the study. We have not yet studied the specific relationship between the variable-range constraints and the strength of the results. We have not examined the relationship between the amount of data in the initial test set, and the value of the final model. We have not studied the relationship between the accuracy of the final model and the effectiveness of the final test set. 

In our ongoing and future work we will seek to explore these questions. We will carry out experiments to examine the effect of variable range on the number of mutants killed. We will look at the accuracy of the inferred model to see if, in this context, it leads to better test sets (building upon the work by Fraser \emph{et al.} \cite{Fraser2015}). We will also investigate the adoption of alternative Machine Learning algorithms that can model more sophisticated types of functionalities, such as complex data structures and sequential behaviour.



\begin{thebibliography}{1}
\bibitem{Hamlet1994}
Hamlet, Richard. "Random testing." Encyclopedia of software Engineering (1994).

\bibitem{Arcuri2012}
Arcuri A, Iqbal MZ, Briand L. Random testing: Theoretical results and practical implications. Software Engineering, IEEE Transactions on. 2012 Mar;38(2):258-77.

\bibitem{Chen2004}
Chen, Tsong Yueh, Hing Leung, and I. K. Mak. "Adaptive random testing." Advances in Computer Science-ASIAN 2004. Higher-Level Decision Making. Springer Berlin Heidelberg, 2005. 320-329.

\bibitem{Claessen2011}
Claessen, Koen, and John Hughes. "QuickCheck: a lightweight tool for random testing of Haskell programs." Acm sigplan notices 46.4 (2011): 53-64.

\bibitem{Goodenough1975}
Goodenough, John B., and Susan L. Gerhart. "Toward a theory of test data selection." Software Engineering, IEEE Transactions on 2 (1975): 156-173.

\bibitem{Raffelt2009}
Raffelt, H., Steffen, B., Berg, T., \& Margaria, T. (2009). LearnLib: a framework for extrapolating behavioral models. International journal on software tools for technology transfer, 11(5), 393-407.

\bibitem{Walkinshaw2009}
Walkinshaw, Neil, John Derrick, and Qiang Guo. "Iterative refinement of reverse-engineered models by model-based testing." FM 2009: Formal Methods. Springer Berlin Heidelberg, 2009. 305-320.

\bibitem{Lei2013}
Feng, Lei, et al. "Case studies in learning-based testing." Testing Software and Systems. Springer Berlin Heidelberg, 2013. 164-179.

\bibitem{Fraser2015}
Fraser, Gordon, and Neil Walkinshaw. "Assessing and generating test sets in terms of behavioural adequacy." Software Testing, Verification and Reliability (2015).

\bibitem{Weyuker1983}
Weyuker, Elaine J. "Assessing test data adequacy through program inference." ACM Transactions on Programming Languages and Systems (TOPLAS) 5.4 (1983): 641-655.

\bibitem{Budd1982}
Budd, Timothy A., and Dana Angluin. "Two notions of correctness and their relation to testing." Acta Informatica 18.1 (1982): 31-45.

\bibitem{Popper1953}
Popper, Karl. "Science: Conjectures and refutations." (1953): pp-33.

\bibitem{Meinke2011}
Meinke, K., \& Sindhu, M. A. (2011). Incremental learning-based testing for reactive systems. In Tests and Proofs (pp. 134-151). Springer Berlin Heidelberg.


\bibitem{Koza1992}
Koza, John R. Genetic programming: on the programming of computers by means of natural selection. Vol. 1. MIT press, 1992.

\bibitem{Seung1992}
Deung, H.S. et al. "Query by Committee." Proceedings of the 5th Annual Workshop on Computational Learning Theory (COLT), 1992

\bibitem{Cherniavsky1987}
Cherniavsky, John C., and Carl H. Smith. "A recursion theoretic approach to program testing." Software Engineering, IEEE Transactions on 7 (1987): 777-784.

\bibitem{Zhu1992}
Zhu, Hong, Patrick Hall, and John May. "Inductive inference and software testing." Software Testing, Verification and Reliability 2.2 (1992): 69-81.

\bibitem{Angluin1987}
Angluin D. Learning regular sets from queries and counterexamples. Information and computation. 1987 Nov 30;75(2):87-106.

\bibitem{Romanik1997}
Romanik, Kathleen. "Approximate testing and its relationship to learning." Theoretical Computer Science 188.1 (1997): 79-99.

\bibitem{Bergadano1996}
Bergadano, Francesco, and Daniele Gunetti. "Testing by means of inductive program learning." ACM Transactions on Software Engineering and Methodology (TOSEM) 5.2 (1996): 119-145.

\bibitem{Valiant1984}
Valiant, Leslie G. "A theory of the learnable." Communications of the ACM 27.11 (1984): 1134-1142.

\bibitem{Briand2009}
Briand, Lionel C., et al. "Using machine learning to refine category-partition test specifications and test suites." Information and Software Technology 51.11 (2009): 1551-1564.

\bibitem{Papadopoulos2015}
Papadopoulos, Petros, and Neil Walkinshaw. "Black-box test generation from inferred models." Realizing Artificial Intelligence Synergies in Software Engineering (RAISE) , 2015.

\bibitem{Ghani2008}
Ghani, Kamran, and John A. Clark. "Strengthening inferred specifications using search based testing." Software Testing Verification and Validation Workshop, 2008. ICSTW'08. IEEE International Conference on. IEEE, 2008.

\bibitem{Vapnik1995}
Vapnik, Vladimir. The nature of statistical learning theory. Springer Science \& Business Media, 1995.

\bibitem{Bishop1995}
Bishop, Christopher M. Neural networks for pattern recognition. Oxford university press, 1995.


\bibitem{Chow1978}
Chow, Tsun S. "Testing software design modeled by finite-state machines." IEEE transactions on software engineering 4.3 (1978): 178.

\bibitem{Settles2010}
Settles, Burr. "Active learning literature survey." University of Wisconsin, Madison 52.55-66 (2010): 11.


\bibitem{Melville2004}
Melville, Prem, and Raymond J. Mooney. "Diverse ensembles for active learning." Proceedings of the twenty-first international conference on Machine learning. ACM, 2004.

\bibitem{Freund1996}
Freund, Yoav, and Robert E. Schapire. "Experiments with a new boosting algorithm." ICML. Vol. 96. 1996.


\bibitem{Poli2008}
Poli, R., Langdon, W. B., McPhee, N. F., \& Koza, J. R. (2008). A field guide to genetic programming. Lulu. com.

\bibitem{Kader1999}
Kader, Gary D. "Means and MADS." Mathematics Teaching in the Middle School 4.6 (1999): 398.

\bibitem{Dietterich2000}
Dietterich TG. Ensemble methods in machine learning. InMultiple classifier systems 2000 Jun 21 (pp. 1-15). Springer Berlin Heidelberg.

\bibitem{Just2011}
Just, R., Schweiggert, F., \& Kapfhammer, G. (2011). "{MAJOR}: An efficient and extensible tool for mutation
		 analysis in a {J}ava compiler" Automated Software Engineering (ASE), 2011.

\bibitem{Jia2011}
Jia, Yue, and Mark Harman. "An analysis and survey of the development of mutation testing." Software Engineering, IEEE Transactions on 37.5 (2011): 649-678.

\bibitem{Arcuri2011}
Andrea Arcuri and Lionel Briand. Adaptive random testing: an illusion of effectiveness?. In Proceedings of the 2011 International Symposium on Software Testing and Analysis (ISSTA '11), 2011.


\end{thebibliography}
\end{document}